\documentclass[%
 reprint,
superscriptaddress,
 amsmath,amssymb,
 aps,
]{revtex4-1}

\usepackage{graphicx}
\usepackage{color} 
\usepackage{dcolumn}
\usepackage{bm}
\usepackage[hidelinks]{hyperref}
\usepackage[utf8]{inputenc}
\usepackage[normalem]{ulem}
\usepackage{amsmath}
\usepackage{siunitx}

\begin{document}

\preprint{APS/123-QED}
    
\title{
An Information-theoretic Collective Variable for Configurational Entropy
}
\author{Ashley Z. Guo}
\thanks{ashley.guo@rutgers.edu}
\affiliation{%
Department of Chemical and Biochemical Engineering, Rutgers University–New Brunswick, Piscataway, New Jersey 08854, USA
}%

\author{Kaelyn Chang}
\affiliation{%
Department of Chemical and Biochemical Engineering, Rutgers University–New Brunswick, Piscataway, New Jersey 08854, USA
}%

\author{Nicholas J. Corrente}
\affiliation{%
Department of Chemical and Biochemical Engineering, Rutgers University–New Brunswick, Piscataway, New Jersey 08854, USA
}%


\begin{abstract}

Entropy governs molecular self-assembly, phase transitions, and material stability, yet remains challenging to quantify and directly control in molecular systems. Here, we demonstrate that the computable information density (CID), a data compression-based information theoretic metric, provides a general per-configuration structural descriptor that tracks configurational entropy changes in molecular dynamics simulations, reflecting both local and long-range structural organization. We validate the CID across systems of increasing complexity, beginning with single-component Lennard-Jones melting before examining binary phase separation, polymer condensation and dispersion, and assembly of amorphous carbon networks at multiple densities. Unlike conventional order parameters, CID requires no \textit{a priori} knowledge of relevant structural features and captures organizational signatures across a variety of molecular systems and discretization resolutions. By establishing a data compression-based structural complexity metric as a practical proxy for configurational entropy, this framework lays a foundation for future entropy-driven materials design and optimization strategies.

\end{abstract}

\maketitle

\section{Introduction}

\begin{figure*}[htbp]
    \centering
    \includegraphics[width=0.85\textwidth]{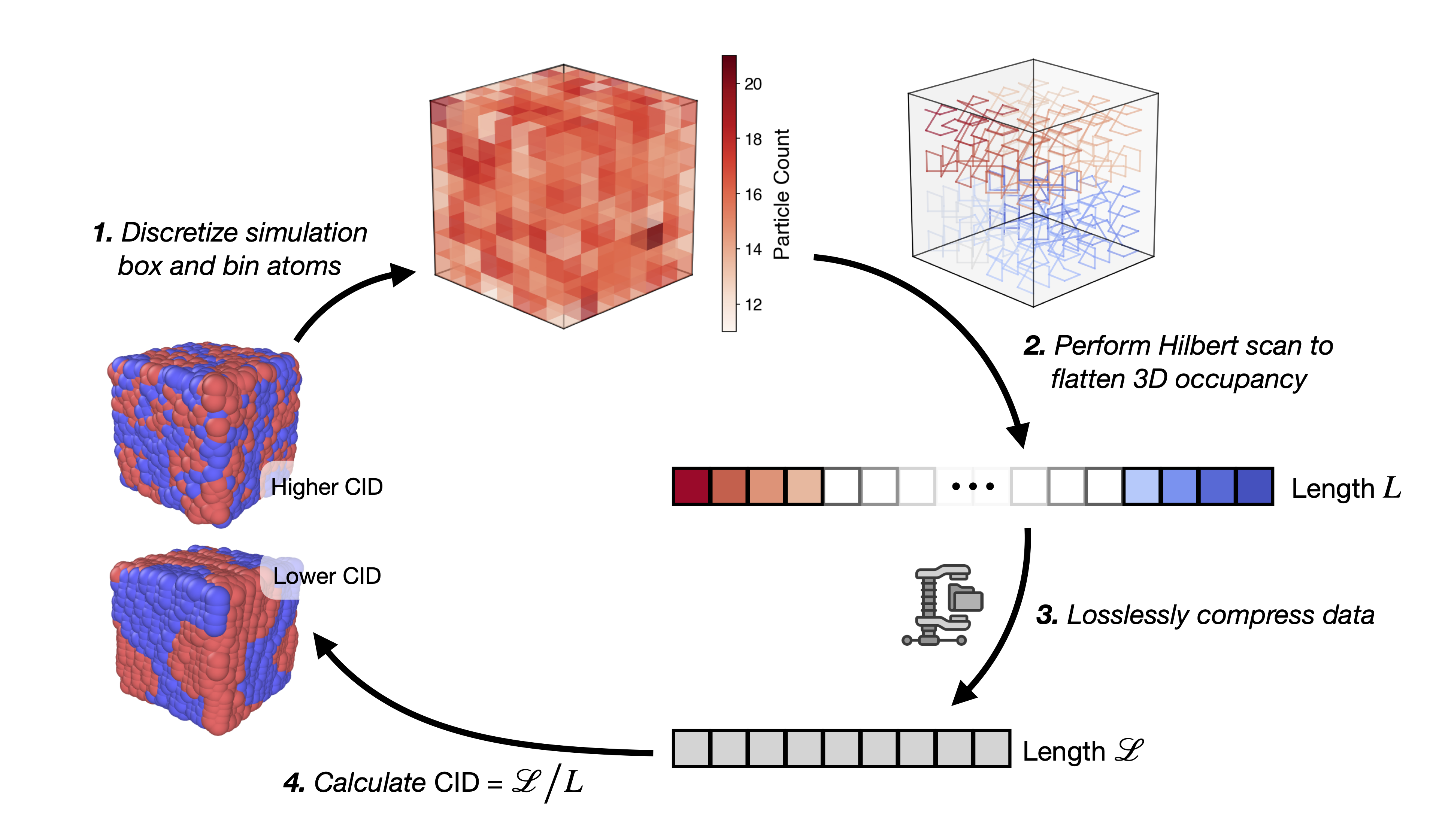}
    \caption{
    Schematic of the CID framework. Atomic coordinates are discretized onto a 3D grid, mapped to a 1D sequence via a Hilbert curve, and compressed using LZ77. CID is calculated as the normalized ratio of compressed to original sequence length, with lower values indicating more ordered configurations.
    }
    \label{fig:schematic}
\end{figure*}

Entropic effects govern molecular behavior across a broad range of scales and contexts. For example, entropy drives protein folding and conformational transitions\cite{Brady1997, Berezovsky2005, Wallin2006, Guo2018hIAPP, Rose2021}, determines phase behavior in multi-component systems\cite{Frenkel1994, Miyazaki2022, Vo2025}, and dictates the stability of self-assembled structures \cite{Mao2013, Harper2019entropicbond, Zarrouk2024}. Despite its central role in molecular thermodynamics, entropy remains far more difficult to directly quantify or engineer than energy. In the context of molecular simulations, potential energies can be directly computed from interatomic forces, and free energies can be calculated from enhanced sampling approaches, making it possible to navigate potential energy landscapes and free energy landscapes through biasing potentials and collective variables (CVs)\cite{Fiorin2013, Sidky2018ssages,Bonomi2009plumed}. However, there currently exists no general descriptor that serves as a proxy for configurational entropy and could enable the identification of entropy-driven transitions or the mapping of entropic contributions across collective variable spaces.

The absence of a general entropy proxy stems from entropy's inherently statistical nature. Unlike energy or structural order parameters that can be evaluated from a molecular configuration's atomic coordinates, entropy has no analogous instantaneous functional form; its statistical mechanical definition $S = -\sum p_i \ln p_i$ requires knowledge of probability distributions over thermodynamic microstates. Existing methods for entropy estimation include quasi-harmonic analysis\cite{Baron2009, Pereira2021}, two-body approximations from pair correlations\cite{Baranyai1990, Sharma2008}, thermodynamic integration\cite{Peter2004}, and machine learning approaches trained on trajectory data\cite{Desgranges2018, Nir2020, BenShimon2025}. These methods can quantify entropy from molecular simulations, but all operate by post-processing ensembles of configurations and impose system-specific assumptions that limit their general applicability. Critically, none provide a per-configuration structural measure that tracks entropic changes and could be rapidly evaluated for individual structures and biased within enhanced sampling protocols. This fundamental gap limits our ability to directly drive molecular systems toward high- or low-entropy states and constrains entropy-based design strategies for materials where configurational entropy governs stability and function.

Foundational work in information theory provides an alternative route to quantifying entropy. Similar to its thermodynamic namesake, Shannon or information entropy $H = -\sum p_i \log_b p_i$ is a measure of the number of possible states of a system or randomness in data, developed in the context of data communication in the 1940s\cite{Shannon1948}. While $p_i$ refers to a probability of being in a specific microstate in the thermodynamic context, $p_i$ in the information context refers to a probability that a certain outcome or event will occur, with units of entropy designated by base $b$ (e.g., base 2 giving units of bits, base $e$ giving units of nats). Shannon's source coding theorem establishes that for a stream of independent and identically distributed random variables, the minimum average number of bits required for lossless compression approaches the information entropy $H$ in the limit of infinite sequence length\cite{Shannon1948}. This theorem establishes a fundamental connection between lossless data compression and entropy: the more compressible data is without information loss, the lower its information entropy. Practical lossless compression algorithms, such as Lempel-Ziv 77 (LZ77\cite{Ziv1977}, used in everyday applications such as zipping files or saving PNG images), approximate this theoretical limit, thereby serving as estimators of information entropy $H$. While this connection between compression and entropy has found applications across diverse scientific domains\cite{Dawy2007, Nalbantoglu2010, Pincus1991, Dyre2024, Zaccone2017}, its use as a proxy for structural order in molecular systems remains underexplored.

Martiniani et al. adapted this compression-as-entropy-quantification principle to quantify order in nonequilibrium systems through the computable information density (CID), defined as the ratio of a compressed sequence of data $\mathcal{L}$ to its original length $L$\cite{Martiniani2019}. CID measures how much a system's spatial configuration can be compressed: highly ordered structures compress more efficiently than disordered ones, directly reflecting the repeating multiscale patterns that characterize more ordered configurations. Crucially, CID requires no \textit{a priori} knowledge of relevant order parameters; it evaluates information content directly from a system's coordinates. This past work demonstrated that CID successfully captures phase transitions in theoretical systems including lattice gases, Manna models, and active Brownian particles, accurately identifying critical points and extracting critical exponents without system-specific tuning. CID has also been applied towards examining quantum phase transitions\cite{chen2025}. Our prior work has also extended this CID analysis to dynamical models of sheared colloids, demonstrating that the CID approach can identify both first- and second-order phase transitions in complex, self-organizing systems\cite{Guo2026}. 

However, these applications of CID are limited to simplified models on discrete lattices or particle systems that lack true thermodynamics. Real atomistic molecular simulations present fundamentally different challenges, including continuous atomic coordinates spanning high-dimensional phase spaces, complex many-body potential energy surfaces with both local and long-range interactions, multiple coupled timescales and lengthscales, and extended molecular structures with specific viscoelastic behavior. Whether data compression-based order quantification can bridge the gap from simpler models to continuous molecular systems, and whether it provides practical advantages over existing entropy calculation methods and order parameters in this regime, remains an open question.

Here, we demonstrate that CID provides a robust, general-purpose structural descriptor that tracks configurational entropy changes across diverse molecular systems and phase transitions. To establish CID as a practical CV, we validate its performance across four criteria: (1) Does it capture changes in order  during phase transitions without system-specific tuning? (2) Can it distinguish thermodynamically distinct states? (3) Does it maintain robustness across different system types and compositions? (4) Is it efficiently computable for practical applications? We address these questions through a series of test systems with increasing complexity, starting with Lennard-Jones models of single-component melting and binary phase separation before moving to temperature-driven polymer condensation and dispersion and the assembly of amorphous carbon networks at varying densities. In each case, CID captures signatures of ordering and disordering without requiring system-specific order parameters, establishing a framework for information-driven characterization of order in molecular simulations.

\section{Methods}

\subsection{Computable Information Density (CID)}

Atomic coordinates from MD snapshots are discretized onto a three-dimensional cubic grid. The user defines a discretization resolution $n$ (constrained to powers of 2, yielding $2^n \times 2^n \times 2^n$ grids) to accommodate later Hilbert curve mapping. In this work, we use $n = 5$ for all systems studied, though we find that $n=4, 5, 6$ work well in each case (Figure~\ref{fig:discretization}, Section III.E). Each grid cell is assigned a character representing its occupancy count: \texttt{0} for empty, \texttt{1} for singly occupied, \texttt{2} for doubly occupied, and so on (using characters \texttt{a}, \texttt{b}, \texttt{c},\dots for occupancies exceeding 9). For dilute single-component systems where multiple occupancy is rare, this effectively reduces to a binary alphabet (\texttt{0}/\texttt{1}). For multi-component systems, occupancy is computed for a user-specified subset of atom types, enabling species-selective structural analysis using the same encoding scheme. The three-dimensional discretized configuration is then mapped to a one-dimensional sequence of length $L$ using a space-filling Hilbert curve\cite{Hilbert1891}. Unlike raster scanning approaches that would only preserve correlations along a single direction, the Hilbert curve maintains spatial locality across all three dimensions: points close in 3D space remain close in the 1D sequence. The power-of-2 grid resolution requirement arises from the recursive construction of the Hilbert curve. The serialized one-dimensional sequence is then compressed using the Lempel-Ziv 77 (LZ77) lossless compression algorithm \cite{Ziv1977}. The algorithm outputs a compressed sequence of length $\mathcal{L}$. The raw CID is calculated as:
$$\text{CID}_{\text{raw}} = \frac{\mathcal{L}}{L}$$ To normalize CID to a more easily physically interpretable scale and account for cells with multiple occupants, we compute a reference CID for a completely disordered system by randomly shuffling the original 1D pre-compression sequence. This shuffle preserves the alphabet (occupancy distribution) while destroying any underlying spatial correlations. The normalized CID is then:
$$\text{CID} = \frac{\text{CID}_{\text{raw}}}{\text{CID}_{\text{shuffle}}}$$

This normalization maps CID to the range [0, 1], where CID = 1 corresponds to a completely random, uncorrelated configuration, and CID $\to$ 0 indicates a uniformly organized configuration. We note that CID$\_$shuffle reflects composition-dependent redundancy while the gap between CID$\_$raw and CID$\_$shuffle captures spatial correlations, a decomposition reminiscent of ideal versus excess thermodynamic properties, though this correspondence is qualitative.

\subsection{Entropy from Radial Distribution Functions}

To validate CID for each example, we compute pair correlation entropy $S_2$ from the radial distribution functions (RDF) $g(r)$:

\begin{equation}
\label{eq1}
S_2/k_B = -2\pi\rho \int_0^{r_{\text{cut}}} [g(r) \ln g(r) - g(r) + 1] r^2 dr
\end{equation}

where $\rho$ is number density and $r_{\text{cut}} = 2.5\sigma$ is the cutoff radius. RDFs were computed at each saved frame using the Freud analysis library \cite{freud2020}, with 200 bins spanning 0 to $r_{\text{cut}}$, followed by numerical integration of Equation~\ref{eq1}.

\subsection{Molecular Dynamics Simulations}

\subsubsection{Lennard-Jones Single-Component Melting}

We simulated a single-component Lennard-Jones fluid consisting of 4000 atoms initialized in an FCC crystal structure at reduced density $\rho^* = \rho\sigma^3 = 0.85$ in a cubic simulation box of side length 10$\sigma$ with periodic boundary conditions. All simulations were performed using LAMMPS \cite{LAMMPS} with reduced units where mass $m = 1.0$, energy $\epsilon = 1.0$, and length $\sigma = 1.0$. Particles interact via the 12-6 Lennard-Jones potential truncated at $r_{\text{cut}} = 2.5\sigma$. The system was initialized at low temperature $T^* = k_BT/\epsilon = 0.01$ and equilibrated for 50,000 timesteps with $\Delta t = 0.001$ using a Nos\'e-Hoover thermostat with damping parameter $0.1\tau$. The crystal was then heated linearly from $T^* = 0.01$ to $T^* = 1.5$ over 500,000 timesteps, driving the solid-to-liquid phase transition. The liquid was then equilibrated at $T^* = 1.5$ for an additional 50,000 timesteps. Atomic coordinates were saved every 500 timesteps throughout the heating trajectory for subsequent CID analysis. CID was computed for the single-component system using $2^5 = 32$ spatial bins per dimension (yielding $32^3$ total grid cells). 

\begin{figure*}[htbp]
    \includegraphics[width=0.85\textwidth]{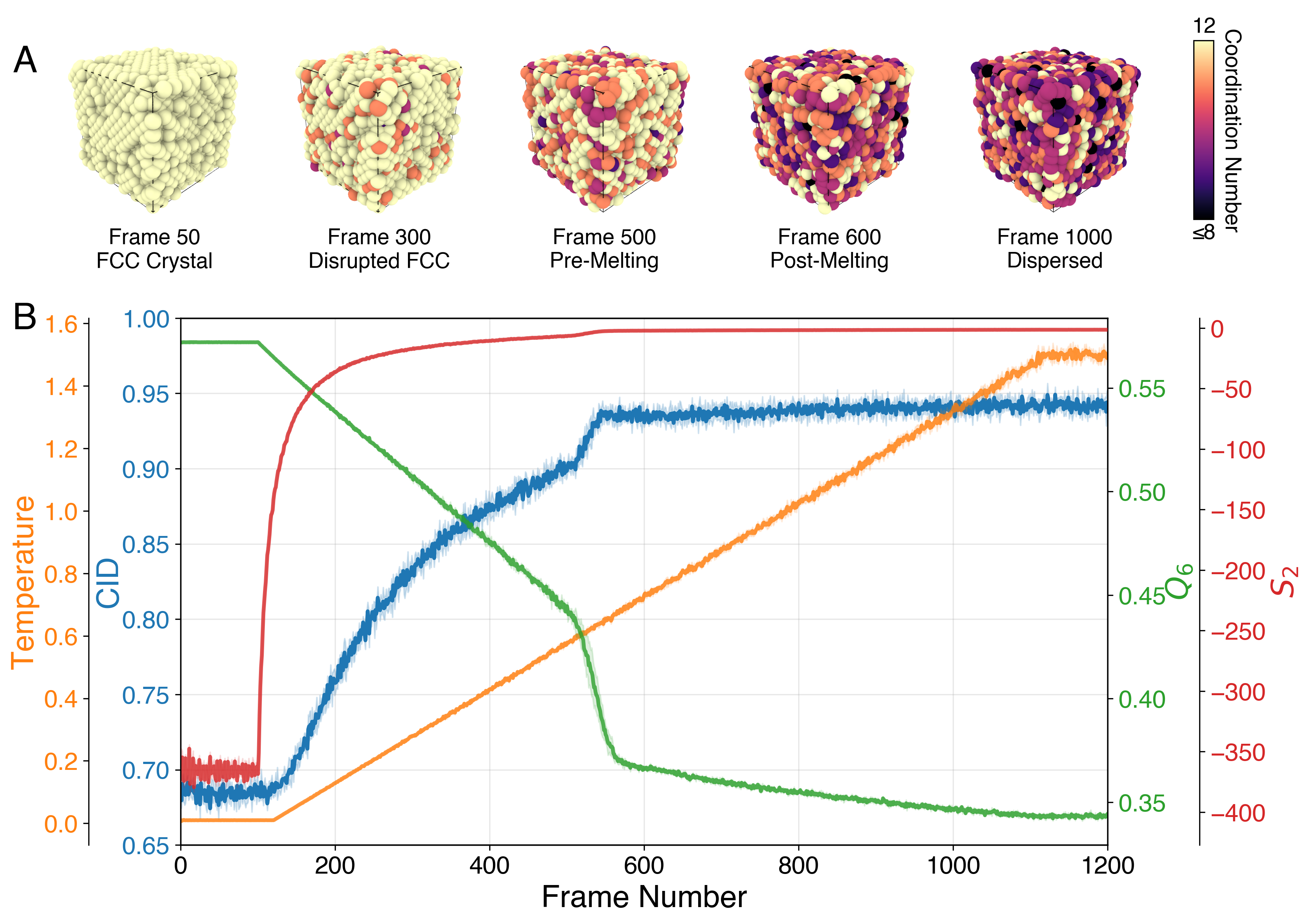}
    \caption{
    Computable information density tracks Lennard-Jones melting transition. (A) Representative snapshots showing FCC crystal (frame 0, $T^* = 0.01$), onset of melting (frame 200, $T^* = 0.4$), mid-transition (frame 400, $T^* = 0.75$), and equilibrated liquid (frame 1200, $T^* = 1.5$). (B) Evolution of CID (blue), temperature (orange), pair correlation entropy $S_2$ (red), and bond-orientational order parameter $Q_6$ (green), during linear heating. CID rises sharply between frames 100-600 as crystalline order breaks down. Both CID and $S_2$ increase during melting while $Q_6$ decreases, but CID shows more gradual evolution reflecting sensitivity to multi-scale structural features. Profiles represent averages over 5 independent simulation runs, with shaded regions indicating one standard deviation.}
    \label{fig:singleLJ}
\end{figure*}

\subsubsection{Lennard-Jones Binary Phase Separation}

We simulated an equimolar binary Lennard-Jones mixture consisting of 13,500 atoms initialized on an FCC lattice at reduced density $\rho^* = 0.8$ in a cubic simulation box with side length $L = 15\sigma$ and periodic boundary conditions. Both species have equal mass $m = 1.0$ and size $\sigma = 1.0$ in reduced units. The system exhibits asymmetric interactions designed to drive phase separation: type A-A interactions have $\epsilon_{AA} = 0.8$, type B-B interactions have $\epsilon_{BB} = 1.2$, and cross-interactions are weakened to $\epsilon_{AB} = 0.2$. All interactions use a Lennard-Jones potential with cutoff $r_{\text{cut}} = 3.0\sigma$ and long-range tail corrections enabled.

The system was initialized with random velocities at $T^* = 5.0$ and equilibrated in the NPT ensemble at $T^* = 5.0$ and $P^* = 1.0$ for 10,000 timesteps ($\Delta t = 0.005\tau$) using a Nosé-Hoover thermostat and barostat with damping parameters of $0.5\tau$ and $5.0\tau$, respectively. Following equilibration, the system was held at $T^* = 5.0$ for 100,000 timesteps, and then the temperature was lowered in a single step to $T^* = 1.0$ and run for an additional 100,000 timesteps. 

CID, $S_2$, and globally averaged Steinhardt bond orientational order parameter $Q_6$\cite{Steinhardt1983} were computed both globally (all atoms) and for each species separately using $2^5 = 32$ spatial bins per dimension. For species-specific metrics, only atoms of the selected type were used for determining occupancy.

\subsubsection{Coarse-grained Homopolymer System}

We simulated a coarse-grained homopolymer melt consisting of 500 chains of 20 beads each in a cubic box of length $L = 40\sigma$ with periodic boundary conditions. Beads interact via a Lennard-Jones potential with $\sigma$ = 1.0 and $\epsilon$ = 1.0 (reduced units), truncated at 2.5$\sigma$. Adjacent beads along each chain are connected by harmonic bonds with equilibrium length $r_0$ = 1.0 and force constant $k$ = 100.0. Chains were randomly initialized and thermalized at $T^* = 5.0$ for 100,000 timesteps ($\Delta t$ = 0.001) using Langevin dynamics to create a well-mixed melt. The system was then cooled from $T^* = 5.0$ to $T^* = 0.1$ over 100,000 timesteps ($\Delta t$ = 0.005) using a Nosé-Hoover thermostat (damping parameter of 1.0$\tau$), followed by equilibration at $T^* = 0.1$ for 100,000 timesteps. We then reheated from $T^* = 0.1$ to $T^* = 5.0$ over 100,000 timesteps, followed by additional equilibration at $T^* = 5.0$. The system was cooled stepwise through $T^*$ = 4.0, 3.0, 2.0, 1.0, and 0.1, with 100,000 timesteps of equilibration at each temperature. Configurations were saved every 1000 timesteps during cooling/heating phases for CID analysis. CID was computed using $2^5 = 32$ spatial binning per dimension. 

\subsubsection{Carbon Systems}

We consider 3D carbon model structures spanning densities from \SI{0.5}{g/cm^3} to \SI{2.0}{g/cm^3}. The carbons were generated using the Annealed Molecular Dynamics (AMD) method\cite{deTomas2017, deTomas2018, CORRENTE2025120160}. Each structure consists of approximately 4000 randomly-distributed carbon atoms in a cubic simulation box with periodic boundary conditions, with box dimensions set to achieve the target carbon density. Carbon interactions are modeled using the reactive EDIP/c potential\cite{Marks2000} which accounts for pairwise, bond, angle, and dihedral interactions and produces physically realistic sp$^2$/sp$^3$ hybridization ratios and graphitization behavior across a range of carbon densities\cite{deTomas2016, deTomas2019}. The structures were annealed at \SI{4000}{K} in the NVT ensemble with a timestep of \SI{1}{fs} for \SI{3}{ns} using a Langevin thermostat. The \SI{0.5}{g/cm^3} and \SI{1.0}{g/cm^3} structures were previously characterized and validated in Ref. \cite{CORRENTE2025120160}.

\begin{figure*}[htbp]
    \includegraphics[width=0.85\textwidth]{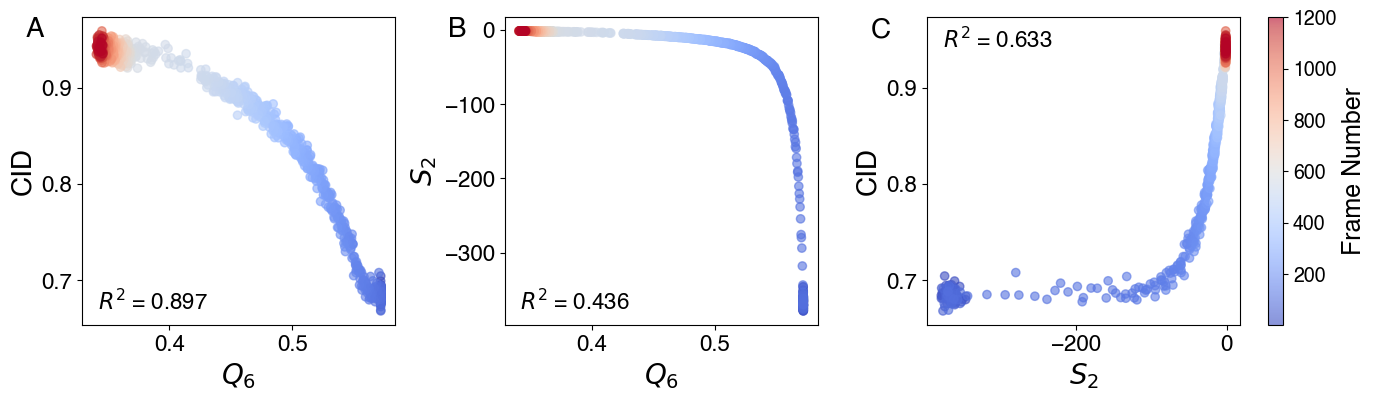}
    \caption{
    Correlations between structural metrics during Lennard-Jones melting (frames 100-600). Scatter plots show relationships between (A) CID vs $Q_6$ ($R^2 = 0.897$), (B) $S_2$ vs $Q_6$ ($R^2 = 0.436$), and (C) CID vs $S_2$ ($R^2 = 0.633$). CID correlates strongly with bond-orientational order while maintaining distinct, nonlinear relationship with pair correlation entropy. Each point represents a single MD snapshot during the melting transition.
    }
    \label{fig:singleLJ_metrics}
\end{figure*}

\section{Results and Discussion}

To establish CID as a practical CV, we must demonstrate four capabilities: (1) capturing entropy changes during phase transitions without system-specific tuning, (2) distinguishing thermodynamically distinct states, (3) maintaining robustness across system types, and (4) computational tractability for enhanced sampling applications.

\subsection{Single-component Lennard-Jones Fluid}

We begin by validating the CID approach on the canonical test case of a single component Lennard-Jones system. Figure~\ref{fig:singleLJ} shows the evolution of CID over the thermal cycle as the system is heated from an FCC crystal solid ($T^* = 0.01$) to liquid ($T^* = 1.5$) over 1200 frames. In the initial crystalline state (frames 0-100, $T^* = 0.01-0.25$), CID remains low ($\approx$0.35), reflecting the highly data-compressible FCC lattice structure. As temperature increases, CID rises sharply between frames 100-600 (T* = 0.25-1.2), tracking the breakdown of crystalline order during melting. The liquid state (frames 600-1200, T* = 1.2-1.5) shows elevated CID ($\approx$0.85), indicating reduced compressibility of the disordered molecular configuration. We note that the data compression gap between raw CID and shuffled CID closes upon melting (Figure S1), consistent with the loss of spatial correlations. 

To validate that CID tracks entropy-correlated structural changes, we compute the pair correlation entropy $S_2$ from radial distribution functions $g(r)$ at each snapshot (Equation~\ref{eq1}). Figure~\ref{fig:singleLJ}B shows the evolution of CID, $S_2$, and the globally averaged Steinhardt bond-orientational order parameter $Q_6$ throughout the melting trajectory. Both CID and $S_2$ increase during the solid-to-liquid transition, with a sharper jump at the onset of melting, consistent with expected behavior for an entropic signature of melting. Simultaneously, $Q_6$ decreases steadily from its expected value of 0.57 for an FCC crystal before sharply dropping at the onset of melting. 

However, CID and $S_2$ exhibit distinct temporal profiles during the transition. $S_2$ increases rapidly early in the melting window (frames 100-300), while CID shows a more gradual rise extending through frame 600. This difference reflects the complementary information captured by each approach: $S_2$ is sensitive to the breakdown of nearest-neighbor spatial correlations encoded in $g(r)$, which begins as soon as particles gain sufficient thermal energy to deviate from lattice positions. CID, through its compression of the full 3D spatial pattern via the Hilbert curve mapping, remains sensitive to longer-range structural correlations and directional order that persist beyond the nearest-neighbor shell. The compression algorithm continues to find exploitable patterns in the partially melted structure even after pair correlations have equilibrated to liquid-like values.

To compare how these different metrics capture the onset of melting, we define the melting window as frames 100--600, spanning the onset of crystalline breakdown (frame 100, where $Q_6$ begins decreasing) to complete liquefaction (frame 600, where all metrics plateau). Within this window, $S_2$ completes 90\% of its total change by frame 195 (19\% through the window), while CID reaches 90\% completion at frame 541 (88\% through) (Figure S2). This difference reflects their distinct sensitivities: $S_2$ rapidly responds to the loss of long-range pair correlations upon void formation, while CID continuously tracks multi-scale structural changes throughout the transition. CID's gradual evolution enables resolution of intermediate structural states during melting (Figure S3), making it well-suited for heterogeneous, multi-stage phase transitions.

CID's strong correlation to $Q_6$ validates its sensitivity to structural transitions while its distinct relationship with $S_2$ shows consistency with conventional entropy estimators (Figure~\ref{fig:singleLJ_metrics}), demonstrating that CID captures information that is complementary to traditional order parameters. The stronger correlation between CID and $Q_6$ compared to $S_2$ reflects a fundamental difference in how these metrics encode structural information. The CID approach preserves spatial locality across all three dimensions, allowing CID to capture directional correlations and orientational order. In contrast, $S_2$ derived from the radial distribution function is inherently spherically averaged and insensitive to angular correlations.

These results establish that the CID satisfies the first requirement for a practical entropy CV, capturing structural changes correlated with thermodynamic entropy during phase transitions without system-specific parameterization. The gradual evolution of CID through the melting window, reflecting its sensitivity to multi-scale structural correlations, suggests it could serve as an effective reaction coordinate for enhanced sampling of melting or crystallization pathways, where traditional order parameters like $Q_6$ may plateau before configurational order fully undergoes transition. 

\subsection{Binary Lennard-Jones Fluid}

To extend CID to multi-component systems, we implement species-selective analysis where occupancy grids are calculated and compression performed separately for each particle type. We validate this approach on a binary Lennard-Jones mixture with asymmetric interactions designed to drive phase separation, with type A atoms having weaker self-interactions while type B atoms have stronger self-interactions, with suppressed cross-interactions. While keeping pressure constant, the system is equilibrated at high temperature before being quenched, inducing demixing (see Methods). 

\begin{figure*}[htbp]
    \centering
    \includegraphics[width=0.85\textwidth]{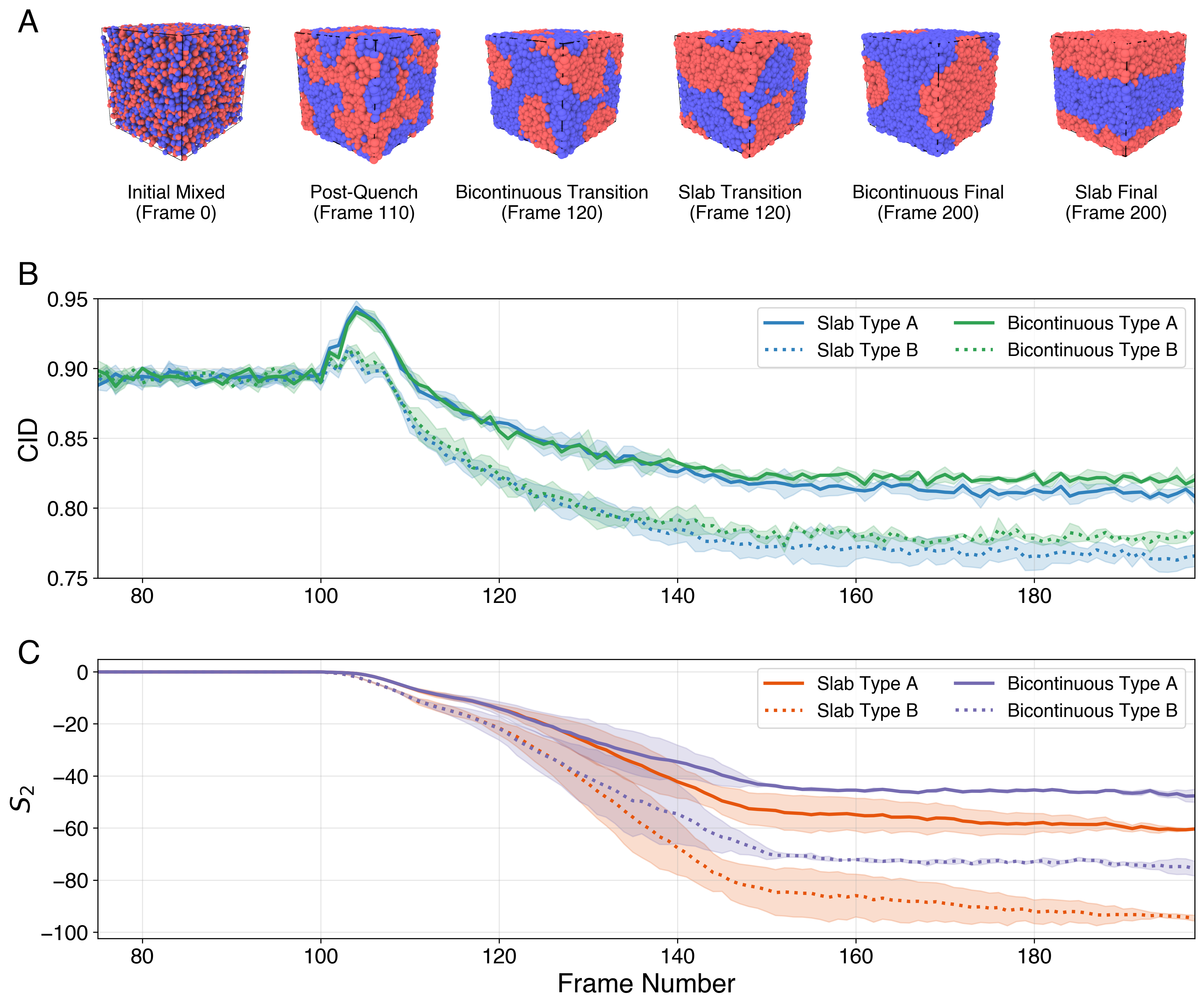}
    \caption{
    CID and $S_2$ track phase separation and distinguish morphologies in binary LJ mixtures. 
    (A) Representative snapshots showing well mixed binary system at high temperature, transient structures, and both slab and bicontinuous phase separated configurations observed at low temperature. 
    (B) CID evolution during equilibration (frames 70-100, T* = 5.0), quench (frame 100), and re-equilibration (frames 100-200, T* = 1.0). The peak in CID near frame 100 reflects NPT volume adjustment. Slab morphologies (blue) converge to lower CID than bicontinuous structures (orange).
    (C) Pair correlation entropy $S_2$ evolution over frames 70-200 shows similar trends to CID but with larger morphology-dependent separation and greater variance within each morphology class, demonstrating CID's complementary sensitivity to global spatial patterns rather than local radial correlations. 
    }
    \label{fig:binaryLJ}
\end{figure*}

Multiple independent simulations produce either slab or bicontinuous morphologies from well-mixed initial conditions (Figure~\ref{fig:binaryLJ}A). Species-selective CID successfully distinguishes particle type based on their self-interaction, with type A atoms consistently exhibiting higher CID values post-quench compared to type B atoms, which is attributed to the stronger self-interactions (and therefore more "ordered" configuration) for type B atoms. CID also detects morphology-dependent differences, where slab geometries show lower CID than bicontinuous structures because the simple spatial partitioning (one half type A, one half type B) is inherently more compressible than interpenetrating networks with more complex interfacial topology. 

Both CID and $S_2$ are able to resolve these features, though $S_2$ shows substantially larger run-to-run variability. This increased uncertainty reflects the sensitivity of $S_2$ to void space when computing pair correlation for spatially segregated species, where radially-averaged $g(r)$ struggles to capture heterogeneity. This limitation becomes more pronounced in systems like polymer condensation, which we explore in Section III.C. 

During the NPT quench, CID exhibits a transient spike (frames 100-110), reflecting density changes as system volume equilibrates in response to the temperature drop, while $S_2$ decreases monotonically. This difference again highlights complementary sensitivities; CID responds to both local density changes and global volume fluctuations, while $S_2$ is primarily influenced by local neighbor orientations. 

The species-selective CID analysis demonstrates extensibility to multi-component systems, a critical requirement for entropy-driven design in alloys, mixtures, and composite materials. Importantly, CID successfully distinguishes both compositional demixing and morphological differences (slab vs. bicontinuous) without requiring \textit{a priori} specification of relevant structural descriptors. The reduced variance in CID compared to $S_2$ for segregated phases further supports CID's utility for driving simulations through inhomogeneous intermediate states where radial correlation functions become ill-defined.

\subsection{Homopolymer Phase Transitions}

To demonstrate CID's applicability to systems with extended, bonded molecular structures, we simulate a coarse-grained homopolymer melt undergoing temperature-driven structural transitions. Unlike the previous Lennard-Jones systems composed of individual non-bonded particles, this system consists of polymer chains with covalent connectivity. We subject the system to a thermal quench from high temperature to $T^* = 0.1$, followed by linear reheating to $T^* = 5.0$, driving the polymer system through a dispersed $\rightarrow$ condensed $\rightarrow$ re-dispersed transition, followed by stepwise cooling back to $T^* = 0.1$ (see Methods).

\begin{figure*}[htbp]
    \centering
    \includegraphics[width=0.85\textwidth]{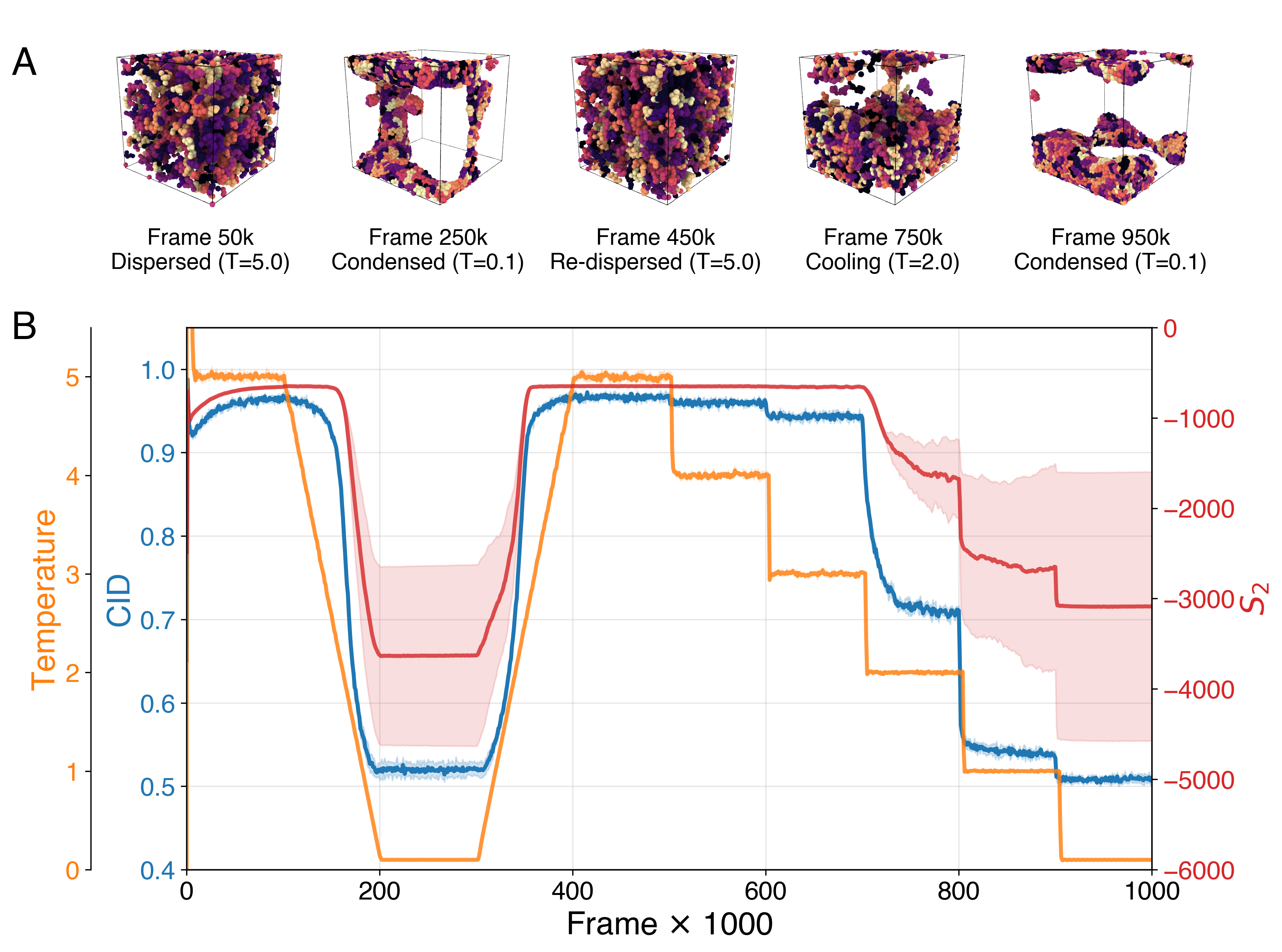}
    \caption{
    Homopolymer structural transitions captured by CID. (A) Simulation snapshots showing dispersed and condensed states at different temperatures with corresponding frame numbers. (B) Normalized CID (blue) and temperature (red) evolution throughout thermal cycling. CID decreases during condensation and increases upon re-dispersion, tracking changes in order without polymer-specific order parameters. 
    Each plotted profile is the mean over 5 independent simulations, with one standard deviation shown in the shaded region. 
    }
    \label{fig:homopolymer}
\end{figure*}

Figure~\ref{fig:homopolymer} shows the evolution of CID throughout the simulation trajectory alongside the temperature profile, pair correlation entropy $S_2$, and Steinhardt parameter $Q_6$. In the initial high-temperature state, the polymer chains are dispersed throughout the simulation box with CID $\approx$ 0.95, indicating a highly disordered spatial configuration that lacks repeating structural motifs. During the temperature quench, CID sharply drops to $\approx$ 0.52 as the system forms a dense, condensed phase. This low CID persists through the low-temperature equilibration regime, reflecting the clear separation between the condensed phase and the surrounding void space, as well as any self-similarity within the condensed phase. Upon reheating, CID rises back to $\approx$ 0.95 as thermal energy drives droplet breakup and chain re-dispersion. During the subsequent stepwise cooling protocol, CID tracks the temperature decreases, showing plateaus at each intermediate temperature before reaching $\approx 0.52$ after the final quench, consistent with the earlier low-temperature region.

While $S_2$ is in qualitative agreement with the CID throughout this temperature protocol, the two metrics are distinctly different in terms of variance. Although the system can condense into very different shapes and morphologies at low temperature, this is handled consistently by the CID, which shows relatively low standard deviation and an average value just above 0.5 in both low-$T$ segments of the temperature ramp. However, $S_2$ does not give a consistent measure in the low-$T$ regime, where its value qualitatively reflects condensation but quantitatively fluctuates over a wide range. This is unsurprising, given that $S_2$ is calculated from $g(r)$ and thus suffers in non-dispersed scenarios, and presents an example where CID is capable of producing a stable measure of order in an inhomogeneous and amorphous system where conventional measures such as $S_2$ are not straightforwardly applicable due to their assumption of spatial homogeneity.

This example highlights a potential advantage of CID for information-driven enhanced sampling in soft matter systems, where it demonstrates stability in precisely the scenarios where conventional approaches fail. The low variance of CID in the condensed regime ($\approx0.52 \pm 0.03$) despite dramatically different morphologies demonstrates robustness to microstates within the same macrostate, a critical property for free energy calculations. The large fluctuations in $S_2$ would pose challenges for use as a biasing coordinate in enhanced sampling, whereas CID's low variance makes it a promising candidate for systems such as protein aggregation, polymer phase separation, or biomolecular condensate formation, where spatial heterogeneity is expected rather than an exception. This motivates future exploration of directly driving assembly or disassembly processes through information-driven biasing potentials.

\subsection{Amorphous Carbons}

\begin{figure*}[htbp]
    \centering
    \includegraphics[width=0.9\textwidth]{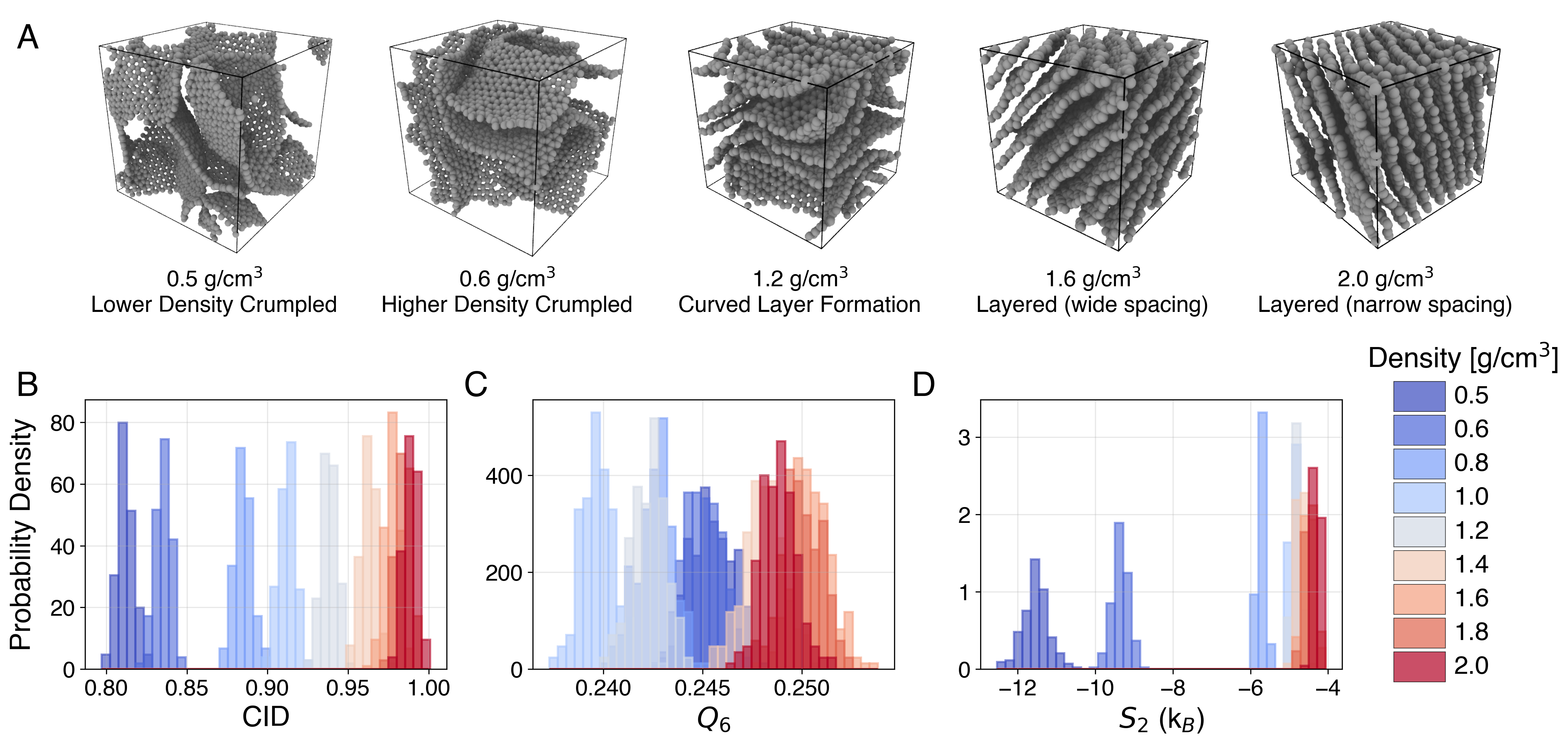}
    \caption{
    Density-dependent structural transitions in amorphous carbon. (A) Representative snapshots show evolution from crumpled networks (0.5-0.6 g/cm³) to ordered graphitic layers (1.6-2.0 g/cm³). (B-D) Distributions of CID, Steinhardt bond orientational order parameter $Q_6$, and two-body entropy ($S_2$) quantify the transition from disordered networks to ordered layered structures.
    }
    \label{fig:carbons}
\end{figure*}

To test CID on more complex, realistic molecular assemblies, we apply it to amorphous carbon structures. These systems exhibit density-dependent structural evolution from disordered graphene networks at low densities, transitioning through higher-density disordered networks, before forming flatter graphitic sheets with well-defined density-dependent inter-layer spacing at higher densities\cite{CORRENTE2025103502}. To assess CID's efficacy in capturing this morphological progression, we again compare it against the pair correlation entropy $S_2$ and $Q_6$ bond orientational order parameter, the latter of which reflects the hexagonal graphitic lattices formed in the carbon structure (Figure~\ref{fig:carbons}).

While $S_2$ monotonically captures the increase in density and is capable of distinguishing low-density structures from layered configurations, it fails to differentiate between layered morphologies at intermediate and higher densities. $Q_6$ at each density is more evenly distributed, though it shows non-monotonic behavior, initially decreasing as disordered, tortuous sheets form into curved sheets before increasing with the emergence of flat graphitic layers. This non-monotonicity, while capturing important symmetry changes, complicates its use as a density-monotonic structural coordinate. CID effectively combines the strengths of both metrics, progressing monotonically through the three structural regimes (crumpled, curved layers, flat sheets) with increasing density. Despite some overlap between adjacent density distributions, CID maintains discriminative power across the full density range, including the high-density regime where $S_2$ saturates. 

\begin{table}[h]
\centering
\caption{Classification accuracy for density prediction from structural metrics in amorphous carbon (5-fold cross-validation, equilibrated frames from $8 \times 10^6$ to $10^7$ timesteps).}
\label{tab:carbon_accuracy}
\begin{tabular}{lc}
\hline
\textbf{Metric(s)} & \textbf{Accuracy} \\
\hline
CID only       & 0.67 $\pm$ 0.01 \\
$Q_6$ only     & 0.30 $\pm$ 0.02 \\
$S_2$ only     & 0.47 $\pm$ 0.02 \\
\hline
$S_2$ + CID    & 0.76 $\pm$ 0.01 \\
$Q_6$ + CID    & 0.74 $\pm$ 0.01 \\
$S_2$ + $Q_6$  & 0.69 $\pm$ 0.02 \\
\hline
All three      & \textbf{0.80 $\pm$ 0.01} \\
\hline
\end{tabular}
\end{table}

To quantitatively assess the discriminative power of each structural metric, we employ Linear Discriminant Analysis (LDA) to predict density class from the computed descriptors, providing a measure of how well each metric (or combination of metrics) captures density-dependent structural changes (Table~\ref{tab:carbon_accuracy}). We report 5-fold cross-validated classification accuracy, where the model is trained on 80\% of the data and tested on the remaining 20\%, repeating this process five times to obtain mean accuracy and standard deviation. While $S_2$ and $Q_6$ alone achieve 47\% and 30\% classification accuracy respectively, their combination reaches 69\%, reflecting their complementary capture of structural ordering and symmetry evolution. In comparison, CID alone achieves 67\% accuracy, and when combined with $S_2$ improves performance to 76\%, indicating it captures structural information orthogonal to the pair-correlation entropy. The combination of all three metrics achieves 80\% accuracy, demonstrating that CID contributes unique structural information not fully captured by traditional metrics. Two-dimensional projections of the order parameter space illustrate the distinct structural information captured by each metric (Figure S4). Notably, CID's monotonic progression across density regimes provides a simpler structural coordinate than the non-monotonic $Q_6$, which may be advantageous for tracking structural evolution in these systems.

We note that despite CID's strong individual performance (67\%), the quasi-2D nature of graphene sheets poses an inherent challenge for CID's 3D voxelization approach: sheet-like carbon structures present primarily as sparse 2D surfaces in 3D space, regardless of their stacking arrangement, suggesting that even higher discriminative power could be achieved with dimensionality-adapted representations. Potential strategies to address this are discussed in the Conclusions.

The carbon network analysis demonstrates CID's appropriateness for studying complex covalently bonded systems that exhibit structural evolution across multiple length scales. CID's strong individual classification accuracy, surpassing both $S_2$ and $Q_6$ alone, indicates that its compression-based representation captures density-dependent morphological changes more effectively than pair correlations or local symmetry measures in isolation. Its combination with conventional metrics achieves considerably improved accuracy, confirming that it also encodes complementary structural information not reducible to these traditional descriptors.  The monotonic progression of CID across density regimes provides a simpler structural coordinate than the non-monotonic $Q_6$, which may potentially be advantageous for optimization algorithms. For materials where target properties correlate with entropy (mechanical stability of low-density carbons, thermal conductivity of graphitic networks), CID provides a structural coordinate for entropy-aware design strategies. More broadly, this example illustrates the importance of representation; for systems where key structural features are not naturally volumetric, hybrid approaches combining CID with topology-aware descriptors may better capture configurational order.

\subsection{Sensitivity to Discretization}

\begin{figure}[htbp]
    \centering
    \includegraphics[width=0.48\textwidth]{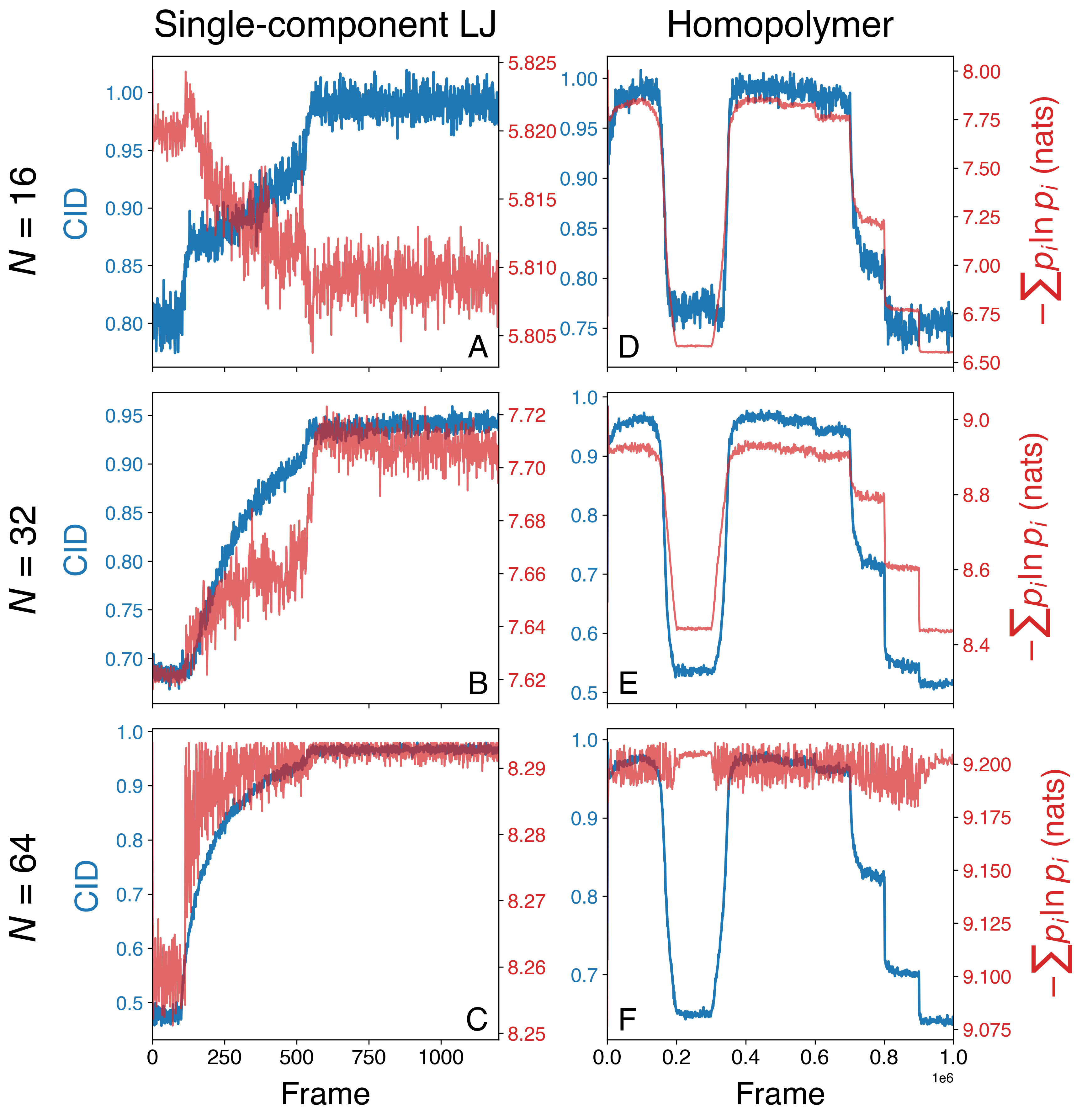}
    \caption{
    Discretization resolution dependence for CID and occupancy-based Shannon entropy. CID (blue) and spatial entropy estimate $-\sum p_i \ln p_i$ calculated from grid occupancy (red) is shown for three different spatial resolutions, for single trajectories of single-component LJ melting (A-C) and homopolymer condensation/dispersion (D-F). Rows show $2^4$ = 16 bins (A,D), $2^5$ = 32 bins (B,E), and $2^6$ = 64 bins (C,F) per dimension. CID maintains qualitatively consistent behavior across all resolutions, capturing phase transitions at both coarse and fine discretizations, with noise increasing as the number of bins decreases. The spatial entropy estimate shows strong dependence on resolution in the inverted behavior at 16 bins (A) and signal loss at 64 bins (C,F).    
    }
    \label{fig:discretization}
\end{figure}

A critical parameter in the CID framework is the spatial discretization resolution, which determines the voxel grid size used to convert continuous atomic coordinates into discrete occupancy patterns. To assess CID's robustness to this choice, we computed CID across three resolutions: $2^4 = 16$, $2^5 = 32$, and $2^6 = 64$ bins per dimension for both single-component Lennard-Jones melting and homopolymer condensation trajectories (Figure~\ref{fig:discretization}).

CID maintains qualitatively consistent behavior across all three resolutions for both systems. In the Lennard-Jones case (Figure~\ref{fig:discretization}A-C), the melting transition is clearly visible at all discretizations, with CID rising from 0.5-0.8 for the crystal (depending on grid size) to 0.95-1.0 for the liquid. The homopolymer condensation trajectory (Figure~\ref{fig:discretization}D-F) similarly shows the dispersion$\rightarrow$condensation$\rightarrow$re-dispersion sequence at all resolutions.

Quantitative CID values show different scaling behavior depending on the system. For the single-component Lennard-Jones system (Figure~\ref{fig:discretization}A-C), CID increases monotonically with resolution: the coarsest discretization ($16^3$) yields the lowest CID values, intermediate resolution ($32^3$) shows higher values, and the finest discretization ($64^3$) produces the highest CID values. This monotonic trend reflects the system's uniform particle size and relatively high density; finer grids capture more detailed spatial features without encountering severe sparsity issues. In contrast, the homopolymer system (Figure~\ref{fig:discretization}D-F) exhibits non-monotonic behavior: in the condensed phase, the CID is lower at the coarsest ($16^3$) and finest ($64^3$) resolutions, with intermediate resolution ($32^3$) yielding the highest CID values. We hypothesize that this reflects the interplay between a dense, relatively uniform core within the condensed phase and its diffuse surface, where coarse discretization fails to resolve surface structure, while fine discretization over-resolves it, creating artificial data sparsity near the interface. 

For comparison, we also computed spatial entropy directly from occupancy probabilities using the log-base Shannon entropy $-\sum p_i \ln p_i$ (Figure~\ref{fig:discretization}, red curves). This ``naive'' entropy estimate shows extreme sensitivity to discretization: in (Figure~\ref{fig:discretization}A), a $16^3$ discretization exhibits inverted behavior, decreasing during melting in contrast with expected entropic trends. At $64^3$ resolution (Figure~\ref{fig:discretization}C,F), the signal in both systems is largely lost. Only at intermediate resolution ($32^3$) does the occupancy-based entropy qualitatively track phase transitions, but even here it shows poor dynamic range and high noise in the case of the single-component LJ system.

This comparison highlights an advantage of CID: the compression algorithm exploits spatial correlations across multiple lengthscales simultaneously, making it robust to discretization choice. While absolute CID values depend on resolution, the metric consistently identifies phase transitions and orders states correctly across a wide range of grid sizes. For the systems studied here, we find that $2^5 = 32$ bins per dimension provides a good balance between capturing molecular-scale structural features and maintaining computational efficiency, which we use for all quantitative analyses in this work.

At the coarsest discretization ($2^4 = 16$), the single-component LJ liquid exhibits normalized CID values slightly exceeding 1.0 (Figure~\ref{fig:discretization}A). This artifact arises from the combination of low atomic density and the uniform spatial distribution of the liquid phase, where the actual configuration produces occupancy patterns that are nearly as random as the shuffled reference used for normalization. Despite this artifact, the CID metric still correctly orders the crystal as more structured than the liquid and accurately captures the melting transition—behavior that stands in stark contrast to the occupancy-based entropy at the same resolution, which shows a completely inverted trend. This demonstrates that even when CID exhibits quantitative artifacts at inappropriate discretization scales, it maintains qualitative correctness, where direct probability-based approaches fail to do so.

The robustness of CID across discretization resolutions ($2^4$ to $2^6$ bins per dimension) confirms its practical utility for enhanced sampling applications, where CV stability is paramount. Unlike ``naive'' spatial entropy, which shows qualitative failures at coarse resolution and signal loss at fine resolution, CID maintains consistent phase transition identification through the compression algorithm's multi-scale pattern recognition. This robustness means users can select discretization based on computational efficiency rather than requiring system-specific optimization. For the intermediate resolution used throughout this work (32 bins), CID calculation adds negligible overhead compared to MD timestep costs, making it computationally tractable even for large-scale enhanced sampling campaigns. The consistent ordering of states across resolutions confirms that CID reflects genuine structural differences rather than artifacts of discretization.

Beyond discretization resolution, the finite length of the symbol sequence also affects CID's discriminative power, as Shannon's source coding theorem strictly applies in the long-sequence limit. We characterize this system-size dependence for our single-species LJ system in Figure S5, where we see the $32^3$ grid used throughout this work provides sufficient sequence length to detect phase transitions, with dynamic range increasing for longer sequences.

\section{Conclusions}

We have demonstrated that computable information density (CID) provides a robust, general-purpose structural descriptor that tracks configurational order in atomistic molecular dynamics simulations. Across systems ranging from simple Lennard-Jones fluids to binary phase separation to polymer melts and amorphous carbon networks, CID successfully captures phase transitions and ordering without requiring system-specific order parameters or collective variables. By establishing a general, per-configuration structural descriptor correlated with entropy, this work addresses a fundamental asymmetry in molecular simulations: while potential energy landscapes and free energy landscapes have long been navigable through geometric optimization and biased sampling, the direct characterization of structural order across configuration space has lacked a comparable general-purpose tool. CID is not a differentiable function of atomic coordinates, making it incompatible with traditional gradient-based enhanced sampling methods. However, its low-cost evaluation as a scalar metric makes it naturally suited for sampling frameworks that require only objective function evaluations rather than analytical gradients.

The CID approach naturally captures spatial patterns across multiple length scales, establishing data compression-based structural characterization as a practical tool for molecular simulations where spatial organization drives multiscale thermodynamic behavior, applicable to a broad class of emergent phenomena central to soft matter and materials science. The method is particularly valuable for systems where traditional order parameters are ambiguous or unknown, enabling study of emergent structural behavior in systems lacking well-defined symmetries or where relevant order parameters are not known \textit{a priori}. We are currently exploring how the CID approach can be applied to investigate structural transitions in MOF systems, where atom-specific CID analysis may reveal subtle rearrangements and breathing or gate-opening behavior.

Beyond the spatial reorganization systems examined here, many molecular systems exhibit entropic transitions that involve both fine-grained orientational or bonding changes and longer scale spatial organization. Examples include secondary structure formation in proteins, coordination environment changes in metal-organic frameworks, bond topology evolution in network-forming materials, and water orientation in hydration shells. For such systems, alternative representations of molecular configurations, where voxel-based spatial information is combined with local orientational descriptors, graph encodings of bond topology, or hybrid approaches, could capture order contributions from multiple structural features and scales simultaneously. The flexibility of data compression-based analysis lies in its representation-agnostic nature: any discrete encoding of configuration space can be compressed, enabling researchers to design representations tailored to specific molecular phenomena while retaining the general framework established here.

Finally, the combination of CID (or its future variations) with learning algorithms promises a new paradigm for computational materials discovery, where entropy is framed as a directly tunable design objective. While traditional approaches to quantifying order rely on physically intuitive descriptors (e.g., symmetries, coordination numbers, bond angles), data compression-based order quantification offers a complementary perspective rooted in information theory. In this view, changes in configurational order are reflected in the data compressibility of a system's discrete representation. By establishing data compressibility as a directly accessible structural metric that tracks entropy-correlated order through this data-centric lens, this framework lays a foundation for future entropy-driven materials design and optimization strategies, with applications to entropy-stabilized materials, assembly pathway optimization, and conformational landscape exploration.

\section*{Author contributions}

\noindent Ashley Z. Guo: Conceptualization, Methodology, Software, Validation, Formal analysis, Investigation, Resources, Data curation, Writing – original draft, Writing – review $\&$ editing, Visualization, Supervision, Project administration. Kaelyn Chang: Investigation, Formal analysis, Data curation, Writing – review $\&$ editing. Nicholas J. Corrente: Investigation, Formal analysis, Data curation, Writing – review $\&$ editing.

\section*{Conflicts of interest}

\noindent There are no conflicts to declare.

\section*{Data availability}

\noindent The code for the computable information density (CID) calculation can be found at \url{https://github.com/guo-group/kappa-py}. All simulation data are available at \url{https://github.com/guo-group/entropic-CID-CV}.

\begin{acknowledgments}

A.Z.G. thanks Gregory Dignon for helpful discussions. 
N.J.C. thanks Nigel Marks of Curtin University for supplying the EDIP/c potential files for LAMMPS. 
The authors acknowledge the Office of Advanced Research Computing (OARC) at Rutgers, The State University of New Jersey, for providing access to the Amarel cluster and associated research computing resources that have contributed to the results here. 

\end{acknowledgments}

\normalem
\bibliography{references}

\end{document}


\preprint{APS/123-QED}

\title{
An Information-theoretic Collective Variable for Configurational Entropy:\\
Supplemental Information
}
\author{Ashley Z. Guo}
\affiliation{%
Department of Chemical and Biochemical Engineering, Rutgers University–New Brunswick, Piscataway, New Jersey 08854, USA
}%

\author{Kaelyn Chang}
\affiliation{%
Department of Chemical and Biochemical Engineering, Rutgers University–New Brunswick, Piscataway, New Jersey 08854, USA
}%

\author{Nicholas J. Corrente}
\affiliation{%
Department of Chemical and Biochemical Engineering, Rutgers University–New Brunswick, Piscataway, New Jersey 08854, USA
}%

\maketitle


\begin{figure}[htbp]
    \centering
    \includegraphics[width=0.65\textwidth]{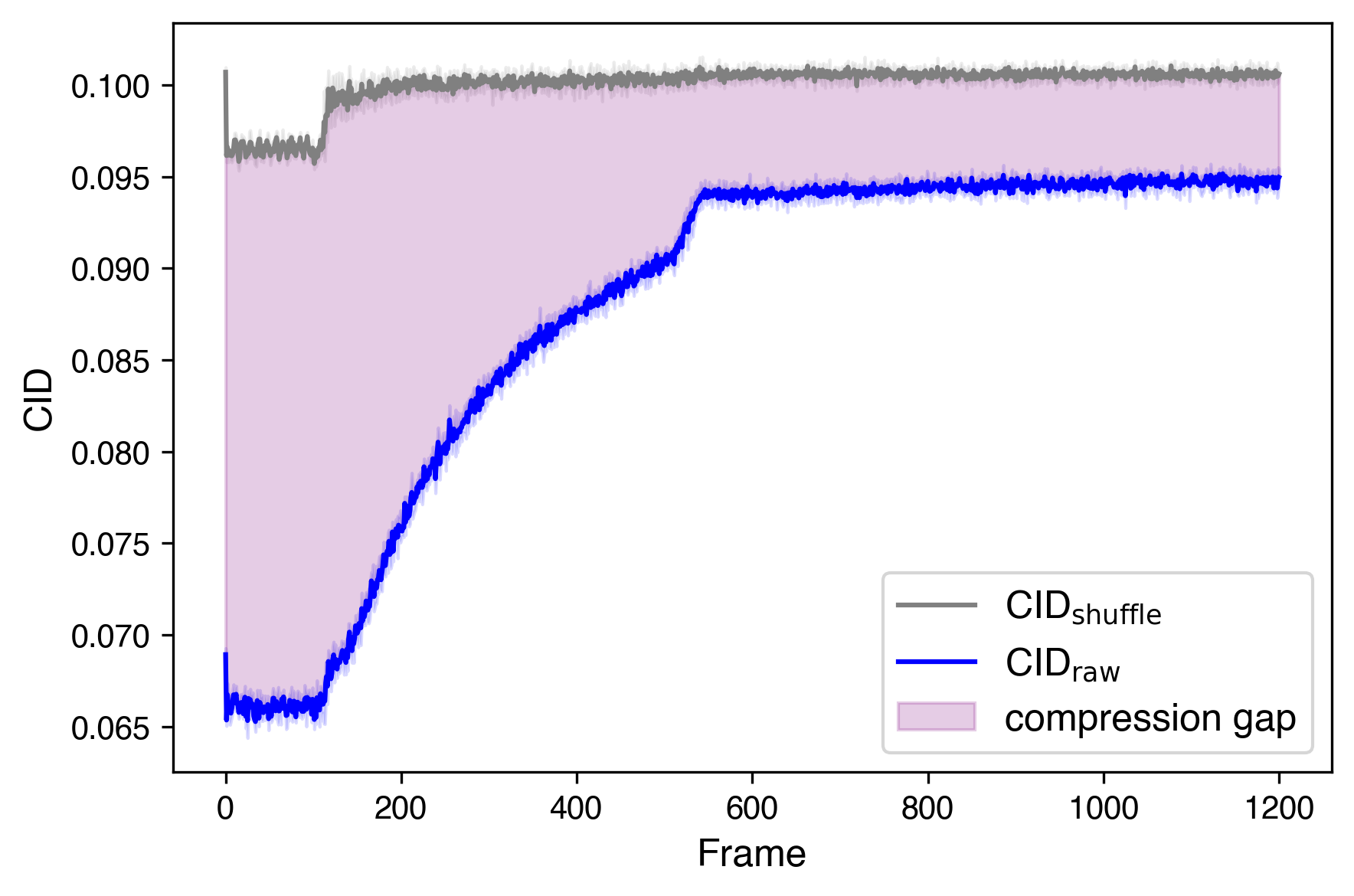}
    \caption{
    Decomposition of CID into composition-dependent and correlation-dependent contributions during single-component LJ melting. CID$\_$shuffle (gray) reflects redundancy from the occupancy distribution alone, while CID$\_$raw (blue) additionally captures spatial correlations. The compression gap is narrowed upon melting, where CID$\_$raw rises sharply as spatial correlations are destroyed, while CID$\_$shuffle increases only slightly from changes in the occupancy distribution. Curves are averaged over 5 independent simulations, with standard deviation shown in shaded regions.
    }
    \label{figsupp:shuffle_gap}
\end{figure}

\clearpage

\subsection{Temporal Sensitivity During Phase Transitions}

\noindent To assess the temporal resolution of different structural metrics, we compare how CID, $S_2$, and $q_6$ evolve during the melting transition of a single-component Lennard-Jones system. We define the transition window as frames 100--600, where the system progresses from the initial defected crystal through complete melting.

\begin{figure}[htbp]
    \centering
    \includegraphics[width=0.85\textwidth]{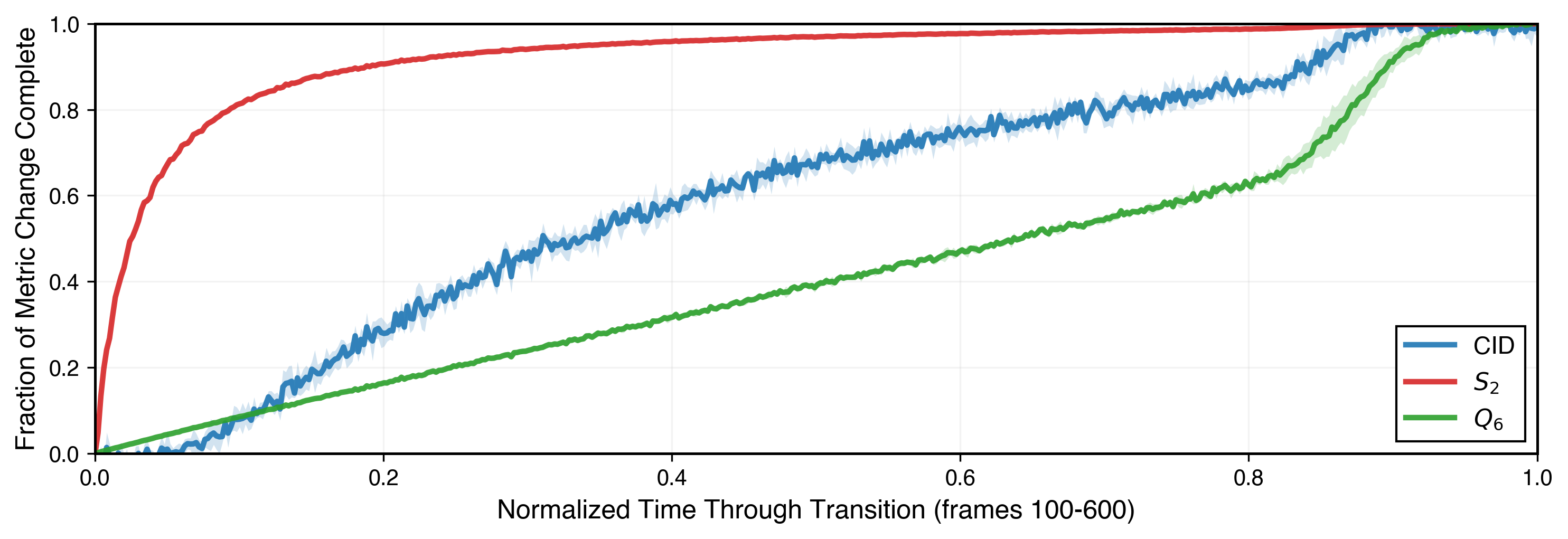}
    \caption{
    Temporal resolution of CID, $S_2$, and $Q_6$ during single-component LJ melting. Each curve shows the fractional completion of the metric's total change through the transition, defined as the window from 100 to 600 frames. $S_2$ rapidly saturates following void formation (\~20\% through transition), consistent with sensitivity to loss of long-range pair correlations. CID continuously captures structural relaxation across the entire melting period. $Q_6$ linearly increases until \~80\% through the observation window, rising sharply at the onset of melting. Curves are averaged over 5 independent simulations, with standard deviation shown in shaded regions.  CID's more gradual response compared to $S_2$ throughout the transition window demonstrates superior temporal resolution for tracking heterogeneous, multi-stage phase transitions.
    }
    \label{figsupp:timing}
\end{figure}

\begin{figure}[htbp]
    \centering
    \includegraphics[width=0.85\textwidth]{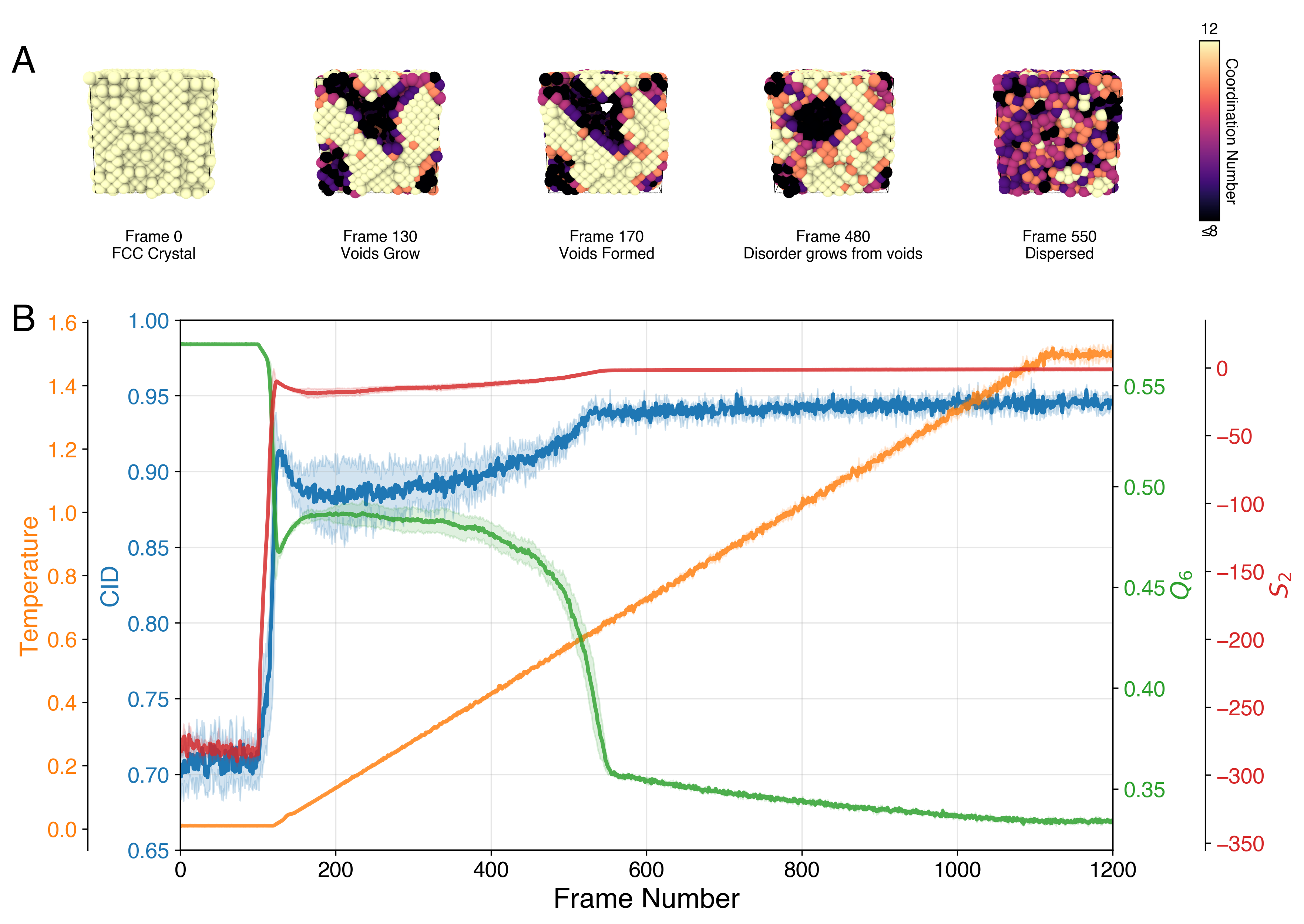}
    \caption{
    Multi-stage melting of single-component LJ fluid at density $\rho^* = 0.8$. (A) Structural snapshots colored by coordination number show void nucleation, growth, and eventual melting. (B) CID (blue) captures multiple order regimes: initial spike from void formation and local defects (frames 100-130), plateau as defected crystal stabilizes (frames 200-400), and gradual rise during thermal disordering and melting (frames 480-550). $S_2$ show qualitatively similar behavior, but with less sensitivity to intermediate states. 
    }
    \label{figsupp:multistage}
\end{figure}

\begin{figure}[htbp]
    \centering
    \includegraphics[width=0.85\textwidth]{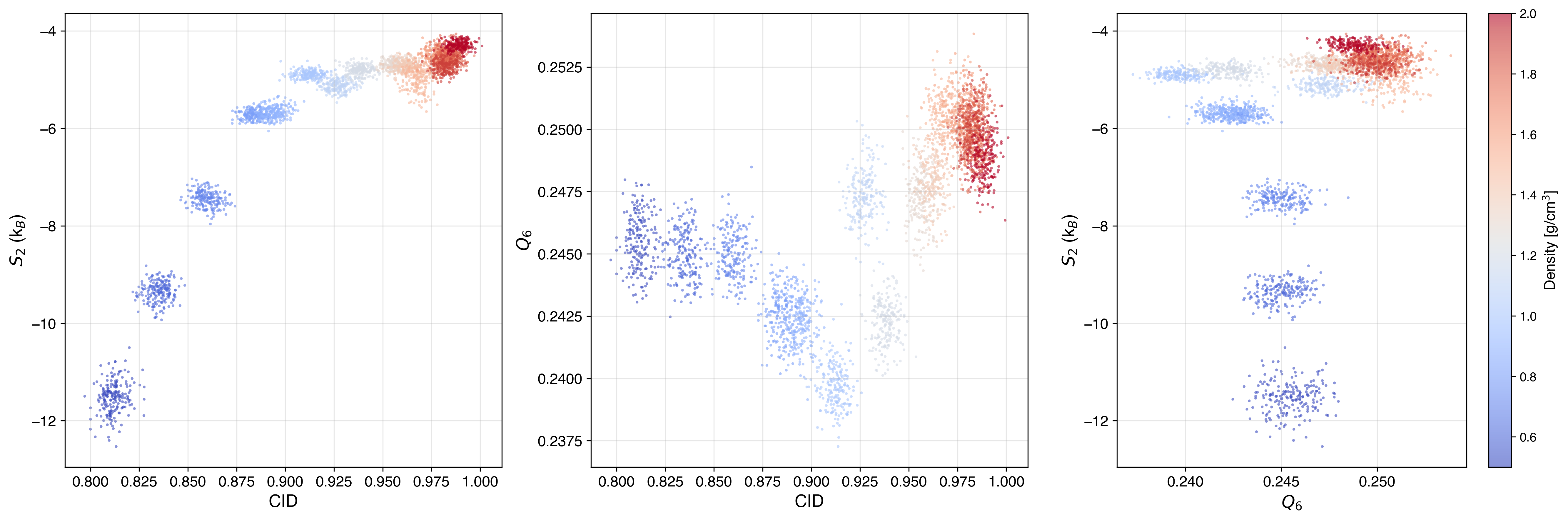}
    \caption{
    Two-dimensional projections of the order parameter space for amorphous carbon structures. Panels show (A) pair correlation entropy $S_2$ vs CID, (B) bond orientational order $Q_6$ vs CID, and (C) $S_2$ vs $Q_6$. Each colored cluster represents configurations sampled at different densities (0.5-2.0 {g/cm$^3$}). $S_2$ saturates at higher densities, failing to distinguish curved from flat layered structures, while $Q_6$ shows non-monotonic behavior, making it unsuitable for tracking order with increasing density. CID provides monotonic progression but with some distribution overlap at higher densities.
    }
    \label{figsupp:carbon_2dcluster}
\end{figure}

\clearpage 

\subsection{System-Size Dependence}

To characterize finite-size effects on CID's discriminative power, we performed three analyses on the single-component LJ melting system. First, the $32^3$ voxel grid from the main text was partitioned into non-overlapping sub-volumes of $8^3$ and $16^3$ voxels, and CID was computed independently on each sub-volume. Second, we simulated a larger system with doubled box length in each dimension at constant density, comparing CID computed on a $64^3$ grid against its non-overlapping $32^3$ sub-volumes. Third, we performed prefix convergence analysis by truncating the Hilbert-curve string at increasing lengths and computing CID for representative crystalline and liquid configurations. All three of these calculations are shown in Figure S5. 

\begin{figure}[htbp]
    \centering
    \includegraphics[width=0.85\textwidth]{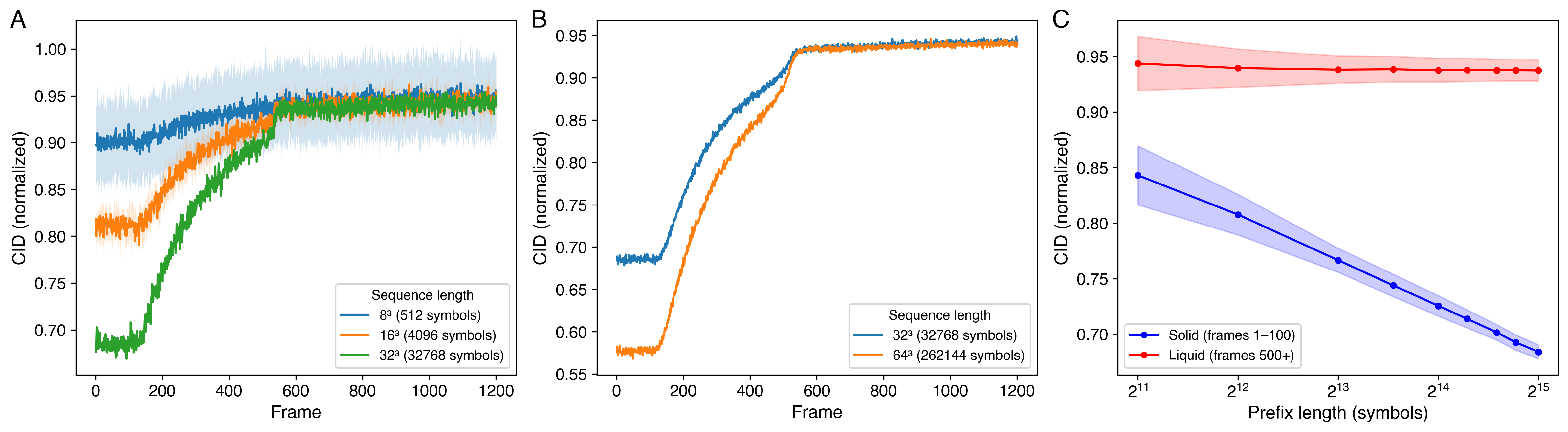}
    \caption{
    Effect of system size and sequence length on CID discriminative power for single-component LJ melting. (A) Sub-volume analysis: the $32^3$ grid is partitioned into non-overlapping sub-volumes of $8^3$, $16^3$, and the original $32^3$ voxels. Dynamic range increases with sequence length; all sizes detect the transition midpoint, though the signal from using $8^3$ voxels is considerably reduced with high variance. Shaded regions indicate one standard deviation over all sub-volumes. (B) Comparison of $64^3$ and $32^3$ grids on a larger system (doubled box length at constant density). (C) Prefix convergence analysis: CID computed on truncated Hilbert-curve strings for representative crystalline and liquid configurations. The liquid converges rapidly while the solid continues decreasing, indicating the compressor exploits increasingly long-range correlations when given additional context. 
    }
    \label{figsupp:size_convergence}
\end{figure}
